\documentclass[twocolumn, switch]{article} %

\usepackage{preprint}
\usepackage{algorithm}
\usepackage{algorithmic}
\usepackage{amsmath, amsthm, amssymb, amsfonts}

\usepackage[numbers,square]{natbib}
\usepackage[utf8]{inputenc}	%
\usepackage[T1]{fontenc}	%
\usepackage{xcolor}		%
\usepackage[colorlinks = true,
            linkcolor = purple,
            urlcolor  = blue,
            citecolor = cyan,
            anchorcolor = black]{hyperref}	%
\usepackage{booktabs} 		%
\usepackage{nicefrac}		%
\usepackage{microtype}		%
\usepackage{lineno}		%
\usepackage{float}			%

\usepackage{multirow}
\usepackage{placeins}
\usepackage{float}
\usepackage{needspace}

\usepackage{newfloat}
\DeclareFloatingEnvironment[name={Supplementary Figure}]{suppfigure}
\usepackage{sidecap}
\sidecaptionvpos{figure}{c}

\usepackage{titlesec}
\titlespacing\section{0pt}{12pt plus 3pt minus 3pt}{1pt plus 1pt minus 1pt}
\titlespacing\subsection{0pt}{10pt plus 3pt minus 3pt}{1pt plus 1pt minus 1pt}
\titlespacing\subsubsection{0pt}{8pt plus 3pt minus 3pt}{1pt plus 1pt minus 1pt}

\title{Multimodal Alignment Improves Generalizability of Genomic Biomarker Prediction in Computational Pathology}

\usepackage{xcolor}
\PassOptionsToPackage{colorlinks=true,linkcolor=gray,urlcolor=gray}{hyperref}

\usepackage{titling}
\usepackage{orcidlink}
\usepackage{footmisc}
\newcommand{\Author}[2]{%
  \textbf{#1}\textsuperscript{#2}\
}

\author{
\Author{Ekaterina Redekop}{1} \and
\Author{Eric Zimmermann}{1} \and
\Author{Ava P Amini}{1} \and
\Author{Alex X Lu}{1} \and
\Author{Neil Tenenholtz}{1} \and
\Author{James Brian Hall}{1} \and
\Author{Lorin Crawford}{1} \and
\Author{Kristen A Severson}{1}
}

\date{%
  \textsuperscript{1}Microsoft Research, Cambridge, MA, United States\\
  [0.5em]
  \footnotesize \textbf{Corresponding author:} \texttt{kseverson@microsoft.com}\\
}

\begin{document}

\twocolumn[ %
  \begin{@twocolumnfalse} %

\maketitle
\thispagestyle{empty}

\begin{abstract}
Computational pathology models that use digitized histopathology whole-slide images have the potential to become a cost-effective and scalable alternative to molecular assays for the prediction of genomic biomarkers, a key task in precision oncology. However, as new genomic biomarkers are discovered or quantified, large, labeled datasets must be prospectively collected to train new models. To address this challenge, we developed MARBLE, a multimodal contrastive pretraining strategy that integrates structured biomarker knowledge into representation learning of histopathology images. MARBLE aligns histopathology-derived representations with representations of genomic biomarkers generated by a large language model (LLM) and a protein language model (PLM). 
This biologically informed alignment enables data-efficient generalization to novel, out-of-distribution biomarkers. 
Using the MSK-IMPACT cohort of over 40,000 patients across multiple biomarker panel versions, we design experiments grounded in real-world data to demonstrate the value of our proposed approach.
\end{abstract}   

\vspace{0.35cm}

  \end{@twocolumnfalse} %
] %

\section{Introduction}
\label{sec:intro}
\begin{figure*}[!t]
    \centering
    \includegraphics[scale=1.1]{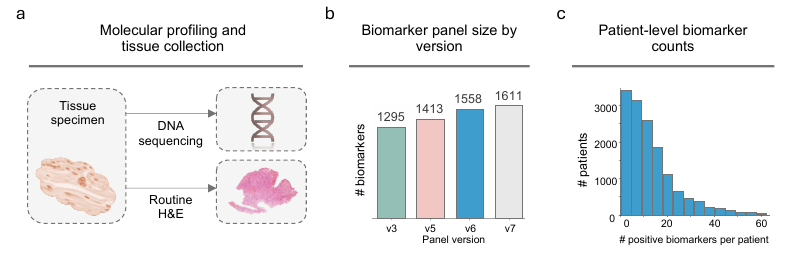} %
    \caption{\textbf{(a)} The molecular profiling and tissue collection pipeline in cancer care: a tissue specimen undergoes DNA sequencing, routine H\&E staining, and digitalization. \textbf{(b)} Biomarker panel size across MSK-IMPACT versions (v3, v5, v6, v7). \textbf{(c)} Distribution of patient-level biomarker counts.}
    \label{fig:main1} %
\end{figure*}

Histopathology data, particularly hematoxylin and eosin (H\&E) stained tissue samples called whole slide images (WSIs), are routinely collected as part of cancer care. These samples are increasingly being digitized, thereby enabling the development and application of foundation models \cite{vorontsov2023virchow, chen2024towards,zimmermann2024virchow2,vaidya2025molecular,juyal2024pluto}, which can be used to extract robust and general-purpose visual representations. Building predictive models using these representations has been successful for routine predictive tasks such as cancer detection and subtyping \cite{ma2024towards, xu2024whole} and has shown great potential for expanding the utility of WSIs to novel downstream tasks. One such use case is the computational prediction of biomarker profiles directly from histopathology images as a replacement for screening for genomic biomarkers via molecular assays (Fig.~\ref{fig:main1}a). DNA sequencing is used to assess the presence of tumor-specific genomic alterations (e.g., mutations, deletions, insertions, duplications, and rearrangements), the results of which are used to inform prognosis and guide therapy selection \cite{cheng2015memorial}. However, as the sequencing procedures underlying these assays are expensive and challenging to scale, direct prediction from H\&E images presents a cost-effective and scalable alternative. Several works have already demonstrated encouraging performance for both single biomarkers \cite{campanella2025real, cifci2022artificial, campanella2025clinical} and larger panels \cite{wang2024screen}.

At the same time, advances in precision oncology continue to expand the catalog of actionable genomic biomarkers, both as new oncogenes are identified and new cancer-driving alterations for existing oncogenes and tumor suppressors are characterized. For example, the MSK-IMPACT panel, a test for tumor-specific genomic biomarkers, has grown by 164 genes since it was first released in 2017 (Fig.~\ref{fig:main1}b)~\cite{cheng2015memorial}. As new cancer drivers and therapeutic targets continue to emerge, predictive models face the challenge of rapidly integrating these biomarkers in a data-efficient manner while maintaining performance on existing ones. In practice, adapting these models to novel targets often requires a large amount of labeled data, which is time-consuming, costly, and potentially infeasible to obtain (e.g., for rare alterations). Addressing this limitation requires approaches that can generalize while utilizing a minimal amount of labeled samples. 

We hypothesize that multimodal contrastive pretraining can overcome this limitation by structuring the visual representation space around biologically meaningful relationships. To this end, we introduce ``Multimodal Alignment foR Biomarker Learning and gEneralization'' (MARBLE), a pretraining strategy that incorporates knowledge about biomarkers directly into the representation learning process, enabling downstream generalization to novel biomarkers (Fig. \ref{fig:main2}). Our approach aligns aggregated H\&E-derived embeddings, created using a pathology foundation model (Path FM), with aggregated semantic embeddings of biomarkers.

To create biomarker embeddings, we use large language models (LLMs) and protein language models (PLMs), allowing us to explore the effect of representing biomarkers as different modalities and to exploit the specific semantics learned by their respective pretraining. 
 Recent studies demonstrate the utility of LLMs in biological domains, from enriching protein representations via textual annotations \cite{duan2024boosting} to modeling perturbations through structured LLM reasoning \cite{wu2025contextualizing}. In parallel, PLMs encode latent information about protein function from amino acid sequences and are thus capable of capturing how mutations impact function \cite{notin2023proteingym, rives2021biological, ye2022progen}. PLMs have been leveraged to incorporate molecular priors into vision models, such as in the VirTues framework \cite{wenckstern2025ai}, which fuses PLM-derived protein embeddings with image patch tokens to encode channel-specific marker information and guide multimodal representation learning in spatial proteomics.

Because tumor morphology often reflects the combined influence of several genomic alterations rather than just one (Fig.~\ref{fig:main1}c),
traditional supervised approaches that treat each biomarker as an independent entity fail to capture the semantic relationships between co-occurring alterations. For example, KRAS, BRAF, and NRAS mutations all activate the MAPK signaling pathway \cite{shirazi2020activating}. By aligning H\&E-derived embeddings with LLM- and PLM-based embeddings, we leverage the structured biological knowledge of these models to capture such relationships. This biologically informed alignment enables the model to learn that morphologically similar tissue patterns may arise from functionally related combinations of biomarkers. From principle, the construction of our method enables generalization to new biomarkers by recognizing their similarity to previously observed alterations with related mechanisms.

We evaluated the MARBLE strategy through experiments grounded in real-world data. Using earlier versions of the MSK-IMPACT panel, we performed pretraining to align image embeddings separately with LLM-derived embeddings and with PLM-derived embeddings, as well as jointly with both. The image-only component of the pretrained model was then fine-tuned to predict all of the biomarkers measured in the most recent version of the panel. Our results support the claim that alignment improves predictive performance for previously unseen biomarkers in a low-data regime and enables generalization to out-of-distribution samples.

Our work makes the following contributions: (1) novel LLM- and PLM-based biomarker representations for use in multimodal contrastive pretraining, (2) a multimodal representation learning framework that incorporates biomarker knowledge into vision features using contrastive learning, evaluated both separately with LLM and PLM embeddings and jointly with both modalities, and (3) a comprehensive set of experiments grounded in real-world settings that demonstrate improved, data-efficient generalization to out-of-distribution biomarkers.

\section{Background and Related Work}

\subsection{Contrastive representation learning in computational pathology} \label{sec:related_wok} 
Contrastive learning aims to align representations of paired samples while simultaneously penalizing the alignment of representations of unpaired samples~\cite{jaiswal2020survey}. This approach has proven highly effective in improving the performance of modern deep learning systems and has recently gained significant interest in multimodal settings such as vision–language modeling. One example is the Contrastive Language–Image Pretraining (CLIP) approach \cite{radford2021learning}, which learns a unified embedding space for images and their corresponding textual descriptions.
Models with multimodal contrastive pretraining are becoming prevalent in computational pathology, spanning both vision–language \cite{shaikovski2024prism, shaikovski2025prism2, ding2024multimodal} and vision–molecular \cite{vaidya2025molecular, jaume2024transcriptomics, redekop2025spade, jaume2024multistain} settings. To date, the use of multimodal pretraining in computational pathology has focused on learning tile aggregation models that produce a slide embedding. In these works, the motivation is that the additional modality provides a generalizable supervisory signal for aggregating tile representations. This is distinct from our setting, where we specifically hypothesize that pretraining will aid in out-of-distribution performance.

\begin{figure*}[!t]
    \centering
    \includegraphics[scale=1]{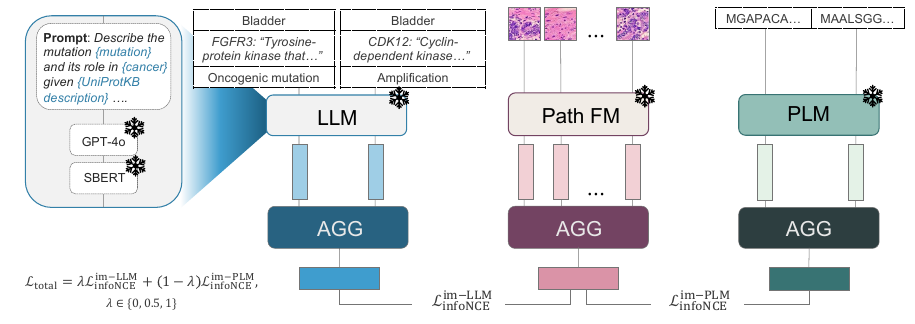} %
    \caption{\textbf{Overview of MARBLE.} Aggregated histopathology embeddings from a frozen pathology foundation model (Path FM) are contrasted with aggregated biomarker embeddings derived from a frozen LLM (GPT-4o + Sentence-BERT (SBERT)) and/or a frozen PLM (ESM-2), enabling cross-modal alignment through contrastive pretraining.}
    \label{fig:main2} %
\end{figure*}

Of the works referenced above, THREADS~\cite{vaidya2025molecular}, a recently introduced slide-level pathology foundation model, is the most closely related to our approach because of the similarity in training data. THREADS was pretrained on 47,171 slides using contrastive learning, where each WSI was paired with matched genomic or transcriptomic profiles. The former were encoded as multi-hot vectors of categorical mutation calls (single nucleotide variants, copy number variants, indels) across 239 genes. Each gene’s status was encoded as a 7-dimensional multi-hot vector, resulting in a final variant feature vector of length 1,673. While this approach captures the presence of a mutation, it discards any biological context about cancer type, oncogene activity, and the mutation itself. In contrast, our approach encodes this information directly into dense biomarker embeddings through LLMs and PLMs. Because our work aims to investigate out-of-distribution generalization and because pretrained checkpoints of THREADS are not available, we re-implement the THREADS approach using our dataset and leverage it as a controlled, comparable baseline. 

\begin{figure*}[!t]
    \centering
    \includegraphics[scale=1]{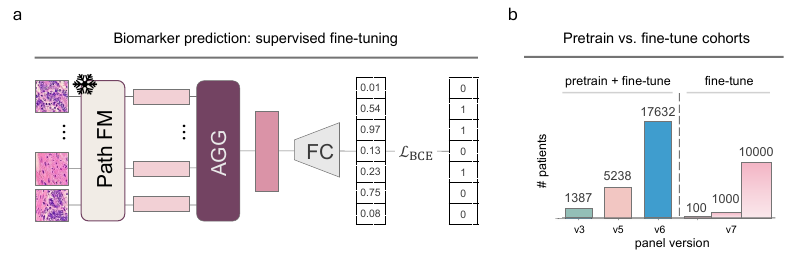} %
    \caption{\textbf{(a)} Supervised fine-tuning of the imaging encoder for multi-label biomarker classification. \textbf{(b)} Pretraining and fine-tuning cohort setups with panel-specific sample counts.}
    \label{fig:main3} %
\end{figure*}

\subsection{Protein language models}
PLMs aim to learn rich representations of protein sequences, capturing the structural and functional properties of proteins through large-scale pretraining~\cite{ferruz2022controllable,ruffolo2024designing}. The recently developed ESM-2 model \cite{lin2023evolutionary} is a particularly noteworthy PLM because of its strong downstream performance in tasks like structure prediction and function annotation. Recently, PLM representations have been combined with imaging data in the context of spatial proteomics to capture both molecular and spatial information with the aim of enabling more accurate clinical diagnostics and biological discovery \cite{wenckstern2025ai}. Building on this idea, we propose to leverage PLM embeddings of mutated protein products of tumor-specific genomic alterations and integrate them with H\&E imaging to derive richer representations that are biologically grounded and predictive of molecular alterations. 

\subsection{Large Language Models}
LLMs are pretrained on massive text corpora via masked language modeling or next token prediction, which implicitly encodes domain-specific knowledge in their weights~\cite{petroni2019language}. Recent studies have demonstrated that this embedded knowledge can be retrieved through prompting. For instance, Med-PaLM, a variant of an LLM tuned to the medical domain, achieved expert-level performance on clinical question-answering tasks using prompt-based querying alone~\cite{singhal2023large}. LLMs are typically used for their ability to directly generate text. In parallel, text embedding models provide compact vector representations of a text corpus, enabling downstream tasks such as multimodal alignment~\cite{reimers2019sentence}. 
\section{Methods}

Our proposed contrastive pretraining strategy \mbox{MARBLE} aligns two types of biomarker representations with histopathology image representations from a patient sample (Fig.~\ref{fig:main2}). 
A foundation model is applied to each modality to produce instance-level embeddings. For histopathology, WSIs are divided into tiles, each embedded independently using a pretrained vision foundation model. For biomarkers, each genomic alteration is represented either as a textual description embedded via an LLM or as an altered protein sequence embedded via a PLM. Both the number of tiles per patient WSIs and the number of biomarkers per patient vary across samples. Therefore, an aggregation step is performed per modality, and the resulting representations are used for contrastive alignment.
After pretraining, the histopathology image-only aggregator is fine-tuned to predict multi-label biomarker status, for biomarkers that were observed and unobserved during pretraining (Fig.~\ref{fig:main3}). Each component is described in further detail below.

\subsection{Embeddings}
Here, we describe the embedding technique used for each of the three modalities. Note that we use $i$ to index the patient sample.

\noindent \textbf{Tile image embeddings.}
WSI crops, commonly referred to as tiles, are embedded using a pathology foundation model, specifically Virchow2~\cite{zimmermann2024virchow2}.
The WSI is split into a grid of high-resolution, non-overlapping tiles extracted at 20$\times$ magnification. Each tile is processed using the foundation model to produce tile-level embeddings. The resulting set of embeddings per patient is denoted
$\textbf{H}_i \in \mathbb{R}^{N_H^i \times d_H}$ where $N_H^i$ is the number of embeddings and $d_H$ is the embedding dimension.

\noindent\textbf{LLM-based biomarker embeddings.}
We propose a systematic approach to transform biomarkers to text embeddings. Each biomarker present in the panel consists of a HUGO gene symbol (e.g., EGFR), high-level mutation information (e.g., copy number amplification), and a cancer type (e.g., bladder cancer).
For each gene, we obtain a canonical protein function description from \mbox{UniProtKB}~\cite{uniprot2018uniprot}, an expert-curated protein knowledge base. The obtained description is then used to prompt an LLM (Fig.~\ref{fig:prompt} for the complete prompt) to produce a concise description of the associated mutation and its role in the provided cancer type, including any known mechanisms of action. The resulting text description is then used as input into a text embedding model, specifically Sentence-BERT (SBERT)~\cite{reimers2019sentence}, to obtain a fixed-dimension embedding $d_L$ per biomarker. This process is repeated for each patient biomarker to create a set of embeddings 
$\textbf{L}_i \in \mathbb{R}^{N_L^i \times d_L}$ where $N_L^i$ is the number of embeddings.

\noindent \textbf{PLM-based biomarker embeddings.} We model in-frame variants of genomic alterations including missense, nonsense, deletion, insertion and indel mutations. Combining the canonical UniProt sequence with the specific genomic alteration provides a sequence of amino acids. These mutated sequences can be used as input to a PLM, providing us a systematic mapping from genomic alteration to embedding. We select ESM-2 as the protein language model \cite{lin2023evolutionary}. This approach produces fixed-dimension embeddings $d_P$ per biomarker which are gathered into a set of $\textbf{P}_i \in \mathbb{R}^{N_P^i \times d_P}$ where $N_P^i$ is the number of embeddings. Copy number variations have no clear altered protein product and are excluded from the PLM representation. Additionally, MSK-IMPACT does not include data about altered protein product from frameshift mutations, so these are also excluded.

\subsection{Aggregation} \label{sec:meth_agg}
Each patient sample contains a variable number of instances, both in terms of tiles and biomarkers. Therefore, all instance-level embeddings must be aggregated into a single vector to enable multimodal alignment at the patient level.

\noindent \textbf{Aggregator for tile embeddings.}
For histopathology tile embeddings, we employ the Agata aggregator module \cite{raciti2023clinical}, which is based on the weakly supervised multiple instance learning paradigm, following previous work \cite{wang2024screen}. The Agata module aggregates a set of tile embeddings $\textbf{H}_i$ %
using a cross-attention mechanism:
\begin{equation}
    \text{Attn}(\textbf{Q},\textbf{K}_{i},\textbf{V}_{i}) = \text{softmax}\Bigg(\frac{\textbf{QK}_{i}^{\top}}{\sqrt{d_{k}}}\Bigg)\textbf{V}_{i}
\end{equation}
where $\textbf{Q}\in\mathbb{R}^{M \times d_{k}}$, $\textbf{K}_{i}\in\mathbb{R}^{N_H^i \times d_{k}}$, and $\textbf{V}_{i} \in \mathbb{R}^{N_H^i \times d_{v}}$ are query, key, and value matrices, respectively, $M$ is the number of learned queries, $d_{k}$ is the dimensionality of the keys, and $d_{v}$ is the dimentionality of the values. Unlike conventional cross-attention, Agata parametrizes $\textbf{Q}$ as a set of learnable query vectors, and keys and values are computed from the tile embeddings $\textbf{H}_{i}$ through learned transformations (Supplementary \ref{sec:supp_agg_emb}).
Attention-weighted embeddings corresponding to different queries are subsequently averaged to obtain a compact patient-level representation
\begin{equation}
    \textbf{h}_{\text{agg},i} = \frac{1}{M}\sum_{m=1}^{M}\text{Attn}(\textbf{Q}_{m},\textbf{K}_{i},\textbf{V}_{i})
\end{equation}
which serves as the input for downstream multimodal alignment.

\noindent \textbf{Aggregator for biomarker embeddings.}
For the sets of biomarker embeddings, $\textbf{L}_i$ and $\textbf{P}_i$, we employ multi-head cross-attention modules, each with a single learnable query vector, $Q\in\mathbb{R}^{1 \times d_{b}}$, that aggregates biomarker-level embeddings into a compact patient-level representation.
Each attention head projects the query and biomarker embeddings into an independent subspace, allowing the model to capture complementary molecular relationships. To handle cases where no biomarkers are present, we introduce a learnable null token. Further details are provided in the Supplementary \ref{sec:supp_agg_biom_emb_llm} and \ref{sec:supp_agg_biom_emb_plm}. The resulting aggregated vectors $\textbf{l}_{\text{agg},i}$ and $\textbf{p}_{\text{agg},i}$ serve as patient-level representations that summarize the set of biomarker embeddings and are used in the downstream multimodal alignment. 

\noindent \textbf{Alternative aggregation strategies.} As an alternative to learnable attention aggregation for the language embeddings, we consider an LLM-based summarization approach. The LLM is first prompted to generate a concise paragraph describing all biomarkers for a given patient (Fig. \ref{fig:prompt_summary} for the complete prompt). This summary text is then embedded using the same sentence-level LLM used for individual biomarkers, resulting in a fixed-length patient-level representation. 

\subsection{Multimodal alignment}
The aggregation models are trained using the symmetric cross-modal contrastive loss that has been widely applied in many recent multimodal pretraining frameworks (e.g., \cite{jaume2024transcriptomics, vaidya2025molecular}). 
Specifically, given a batch of $N$ image-LLM pairs, $\{(\textbf{h}_{\text{agg},i}, \textbf{l}_{\text{agg}, i})\}_{i=1}^{N}$,
we first project the representation for each modality into a common $d$-dimensional space using modality-specific projectors,
followed by $L_2$ normalization to produce $\hat{\textbf{h}}_i$ and $\hat{\textbf{l}}_i$.  
We then optimize the symmetric, normalized, and temperature-scaled contrastive loss %
\begin{equation}
\begin{split}
\mathcal{L}_{\mathrm{con}}
&= -\frac{1}{N}\sum_{i=1}^{N}
\log\frac{\exp(\hat{\textbf{h}}_i^\top \hat{\textbf{l}}_i/\tau) }{\sum_{j=1}^{N}\exp(\hat{\textbf{h}}_i^\top \hat{\textbf{l}}_j/\tau)}
 \\
&-\frac{1}{N}\sum_{i=1}^{N}
\log\frac{\exp(\hat{\textbf{l}}_i^\top \hat{\textbf{h}}_i/\tau)}{\sum_{j=1}^{N}\exp(\hat{\textbf{l}}_i^\top \hat{\textbf{h}}_j/\tau)}
\end{split}
\end{equation}
where $\tau$ is a learnable temperature parameter that scales the similarity logits. An analogous approach is used for sets of image-PLM pairs $\{(\textbf{h}_{\text{agg},i}, \textbf{p}_{\text{agg}, i})\}_{i=1}^{N}$. 
For experiments where two-sided alignment is performed between the aggregated tile representation and aggregated LLM- and PLM-based representations, the total loss is computed as a weighted average of the two symmetric contrastive terms:
\begin{align}
\mathcal{L}_{\text{total}} 
= \lambda \, \mathcal{L}_{\mathrm{con}}^{\text{image--LLM}}
+ (1-\lambda)\, \mathcal{L}_{\mathrm{con}}^{\text{image--PLM}},
\end{align}
where $\lambda \in \{0,0.5,1\}$ balances the contribution of the two alignment objectives and unifies all model variants: MARBLE-LLM corresponds to $\lambda=1$ (alignment with LLM embeddings only), MARBLE-PLM corresponds to $\lambda=0$ (alignment with PLM embeddings only), and \mbox{MARBLE} corresponds to $\lambda=0.5$ (joint alignment with both modalities).

\subsection{Biomarkers prediction}
After pretraining the multimodal alignment module, we initialize the biomarker classification model with the image aggregator and projection head weights, and we jointly fine-tune them with a multilabel classification head to predict multiple biomarkers simultaneously using a binary cross-entropy loss $\mathcal{L}_{BCE}$ (Fig.~\ref{fig:main3}a).

\section{Experiments}
We design our experiments and evaluations to mirror a real-world setting. 
Using the MSK-IMPACT panel genomic alterations as our target predictive task, we exploit the property that several versions of the panel have been released over time, each adding new biomarkers. This allows us to test the effect of pretraining with early versions and fine-tuning with varying numbers of labeled samples from the latest panel.
Details are provided below. 
\subsection{Dataset}
\label{sec:dataset}
The cohort used in this study underwent paired tumor-normal targeted sequencing with the FDA-approved \mbox{MSK-IMPACT} assay at Memorial Sloan Kettering Cancer Center. Each assay sample is paired with one or more H\&E-stained WSIs obtained from the same formalin-fixed paraffin-embedded tissue block. Each biomarker is defined as a pair: a gene name and a genomic alteration. Genomic alterations are defined as one of the following: mutations (single-nucleotide variants (SNVs) and insertions/deletions (indels)), copy number variations (amplifications and deletions), and structural rearrangements (fusions). 

The cohort contains 40,300 patients tested with one of four \mbox{MSK-IMPACT} panel versions (v3, v5, v6, or v7); note that a v4 panel was never released. The number of patients tested with a specific version, as well as the number of oncogenes and biomarkers associated with this version, are shown in Table \ref{tab1}.
During pretraining, only biomarkers from panels v3, v5, and v6 are used. This is critical to ensure no data leakage and test true out-of-distribution generalization.
For supervised fine-tuning, in all experiments, 2681 patients with biomarkers measured using the v7 panel (all biomarkers) are used as the test set. An additional 2692 patients with biomarkers measured using the v7 panel are used as the validation set during supervised fine-tuning.

\subsection{MARBLE setup}
Based on data availability, we considered two pretraining scenarios.

\noindent \textbf{Scenario 1: LLM only.}
Pretraining was performed using all samples measured with versions v3, v5, and v6, spanning 24,257 patients in total. MARBLE-LLM, the multi-hot baseline described in Sec.~\ref{sec:MH}, the alternative aggregation approach described in Sec.~\ref{sec:meth_agg}, and a supervised baseline were evaluated.

\noindent \textbf{Scenario 2: All settings.}
Protein mutation data were obtained from the publicly available MSK-IMPACT 2017 cohort through cBioPortal \cite{zehir2017mutational}. Detailed annotations specifying the exact amino acid substitutions were available only for the two earlier panel versions (v3 and v5). This data was essential for constructing altered protein sequences, which were subsequently embedded with a PLM to capture the molecular changes in the embedding space. This same subset (v3 and v5) of patient samples was also embedded with the LLM approach.
Consequently, pretraining using MARBLE-LLM, MARBLE-PLM, and MARBLE was performed on this subset.

\begin{table}[]
\small
\resizebox{0.47\textwidth}{!}{
{\renewcommand{\arraystretch}{1} 
\begin{tabular}{lcccc}
\toprule
Panel Version & v3   & v5   & v6    & v7    \\ \hline
Patients      & 1387 & 5238 & 17632 & 16043 \\
Oncogenes         & 341  & 410  & 468   & 505 \\ 
Biomarkers    & 1295 & 1413 & 1558  & 1611  \\
Sequence available & $\checkmark$ & $\checkmark$ & &  \\ \bottomrule
\end{tabular}}}
\caption{\textbf{Summary of patient cohort size and number of biomarkers for MSK-IMPACT panel versions (v3, v5, v6, v7).}}
\label{tab1}
\end{table}

\begin{figure*}[!t]
    \centering
    \includegraphics[scale=0.9]{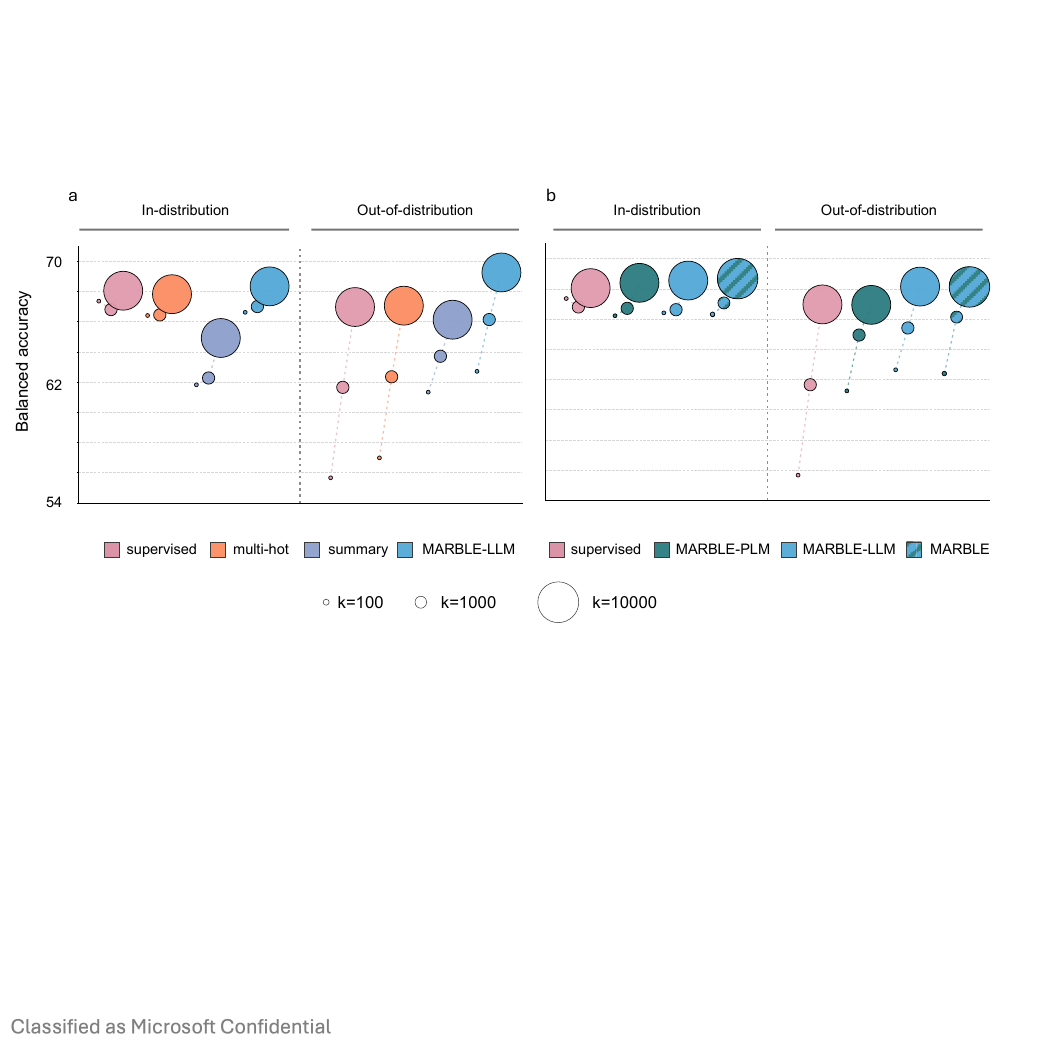} 
    \caption{\textbf{Performance comparison of multimodal pretraining across generalization setups and data regimes}. Balanced accuracy (BA) of in-distribution and out-of-distribution biomarker prediction at three data regimes ($k$ = 100, 1000, 10000).
    (\textbf{a}) Results for Pretraining Scenario 1, using v3, v5, and v6 panels.
    (\textbf{b}) Results for Pretraining Scenario 2 using v3 and v5 panels.}
    \label{fig:res_bubbles} 
\end{figure*}

\noindent \textbf{Supervised fine-tuning.}
For both Scenarios, during fine-tuning, the pretraining data from Scenario 1 is combined with a random sample of $k$=100, 1000, or 10,000 patients measured with the v7 panel to train the multi-label classification model for all 1611 biomarkers (Fig.~\ref{fig:main3}b). This procedure is repeated five times to ensure robustness. Although the samples are random, they are consistent across settings to create valid comparisons. For all pretrained settings, weights are initialized with the corresponding MARBLE or baseline model. In the supervised baseline, the architecture and training data are the same, but the weights are initialized randomly.

We measured performance using the average of per-biomarker balanced accuracy (BA), with stratification for in-distribution, out-of-distribution and all biomarkers. In-distribution refers to biomarkers included during pretraining (i.e., up to v6 of the panel). Out-of-distribution refers to biomarkers not included during pretraining and introduced only at the fine-tuning stage (i.e., the latest panel, v7). To calculate balanced accuracy, the binarization threshold for each biomarker was selected based on the validation subset using the Youden index criterion \cite{youden1950index} to maximize sensitivity–specificity trade-off.
The reported values represent the mean and standard deviation across the 5 trials.

\subsection{Comparison approaches} \label{sec:MH}
To show the benefit of LLM- and PLM-based biomarker representations, we include a multi-hot baseline following the previous work by Vaidya et al. \cite{vaidya2025molecular} (refer to Sec.~\ref{sec:related_wok} for additional detail). Each patient's biomarker panel is represented by a binary vector of length equal to the total number of unique biomarkers in the dataset, where each element indicates the presence (1) or absence (0) of a specific biomarker.
This multi-hot vector is then passed through a shallow feed-forward network to project it into the same latent dimensionality as the other biomarker encoders, ensuring comparability during multimodal alignment. Contrastive pretraining and supervised fine-tuning proceed in the same manner as the other experiments.

\subsection{Implementation details}
Contrastive pretraining and supervised fine-tuning both use the AdamW optimizer \cite{loshchilov2017decoupled} using an initial learning rate of $10^{-4}$ and weight decay 
of 0.05. The learning rate follows a cosine annealing schedule, with a maximum of 50 epochs for contrastive pretraining and 20 epochs for fine-tuning. During pretraining, early stopping is applied based on the validation loss of in-distribution biomarkers up to and including the v6 panel, while during supervised fine-tuning, early stopping is based on the validation performance measured by the AUC on all biomarkers up to and including the v7 panel, including out-of-distribution ones. A global batch size of 64 was used for both pretraining and fine-tuning. Additional implementation details are provided in \mbox{Section \ref{sec:suppl_impl_det}}.

\section{Results}

\begin{table*}[!tp]
\centering
\small
{\renewcommand{\arraystretch}{1.15} 
\begin{tabular}{lccc|ccc|ccc}
\toprule
           & \multicolumn{3}{c|}{In-distribution}                                                                                               & \multicolumn{3}{c|}{Out-of-distribution}        & \multicolumn{3}{c}{Overall}                                                                                            \\ \hline
           & k=100                               & k=1000                                       & k=10000                                       & k=100                                        & k=1000                                       & k=10000    & k=100 & k=1000 & k=10000                                  \\ \hline
Supervised & \textbf{67.4 ± 0.4}            & 66.8 ± 0.3                              & 68.1 ± 0.1                            & 55.7 ± 0.6                              & 61.7 ± 1.3                              & 67.0 ± 0.9      &  67.0 ± 0.4 & 66.6 ± 0.4 & 68.0 ± 0.2                        \\
Multi-hot    & 66.4 ± 0.3                     & 66.5 ± 0.3                             & 67.8 ± 0.2                               & 57.0 ± 1.1                              & 62.4 ± 1.1                              & 67.1 ± 0.7       & 66.1 ± 0.3 & 66.3 ± 0.3 &      67.8 ± 0.2                 \\
Summary    & 61.8 ± 0.2                     & 62.3 ± 0.2                              & 64.9 ± 0.2                               & 61.3 ± 1.4                              & 63.7 ± 0.5                              & 66.1 ± 0.6              &  61.8 ± 0.3 & 62.3 ± 0.3 & 65.0 ± 0.2                \\
MARBLE-LLM        & \multicolumn{1}{l}{66.6 ± 0.3} & \multicolumn{1}{l}{\textbf{67.0 ± 0.2}} & \multicolumn{1}{l|}{\textbf{68.4 ± 0.2}} & \multicolumn{1}{l}{\textbf{62.7 ± 0.8}} & \multicolumn{1}{l}{\textbf{66.2 ± 1.0}} & \multicolumn{1}{l|}{\textbf{69.3 ± 1.5}} & \textbf{67.1 ± 0.4} & \textbf{66.7 ± 0.3} & \textbf{68.5 ± 0.2} \\ \bottomrule
\end{tabular}
}
\caption{\textbf{Balanced accuracy results comparing MARBLE-LLM with supervised and multi-hot baselines, as well as an LLM summary ablation.} Performance is reported for in-distribution biomarkers (present during pretraining from earlier panel versions, $n = 1558$), out-of-distribution biomarkers (introduced in the latest panel, $n = 53$), and the combined overall set ($n = 1611$). During fine-tuning, $k$ = 100, 1000, and 10,000 patients were sampled from the latest panel version (v7). Each value represents the mean ± standard deviation over five independent random draws of $k$.}
\label{tab:res1}
\end{table*}

\begin{table*}[!tp]
\centering
\small
{\renewcommand{\arraystretch}{1.15} 
\begin{tabular}{lccc|ccc|lll}
     \toprule
           &  \multicolumn{3}{c|}{In-distribution}                      & \multicolumn{3}{c|}{Out-of-distribution}                  & \multicolumn{3}{c}{Overall}                                                          \\ \hline
     & k=100             & k=1000            & k=10000           & k=100             & k=1000            & k=10000           & \multicolumn{1}{c}{k=100} & \multicolumn{1}{c}{k=1000} & \multicolumn{1}{c}{k=10000} \\ \hline
MARBLE-PLM  & 66.2 ± 0.2          & \textbf{66.7 ± 0.2}          & 68.4 ± 0.3          & 61.3 ± 0.9          & 65.0 ± 0.3          & 67.0 ± 0.4          & 66.2 ± 0.1                & \textbf{66.7 ± 0.2}       & 68.5 ± 0.3                  \\
MARBLE-LLM  & \textbf{66.4 ± 0.3} & 66.6 ± 0.3          & 68.6 ± 0.1          & \textbf{62.7 ± 0.5} & 65.4 ± 0.6          & \textbf{68.2 ± 0.2} & \textbf{66.4 ± 0.3}       & 66.6 ± 0.2                 & 68.4 ± 0.2                  \\
MARBLE & 66.1 ± 0.3        & 66.6 ± 0.1 & \textbf{68.7 ± 0.2} & 62.4 ± 1.8          & \textbf{66.2 ± 0.5} & \textbf{68.2 ± 0.6} & 65.9 ± 0.3                & 66.6 ± 0.1        & \textbf{68.6 ± 0.2}         \\ 
\bottomrule
\end{tabular}

}
\caption{\textbf{Comparison of MARBLE-LLM, MARBLE-PLM and MARBLE, reported in terms of balanced accuracy.} Performance is reported for in-distribution biomarkers (present during pretraining from earlier panel versions, $n = 1558$), out-of-distribution biomarkers (introduced in the latest panel, $n = 53$), and the combined overall set ($n = 1611$). During fine-tuning, k = 100, 1000, and 10,000 patients were sampled from the latest panel version (v7). Each value represents the mean ± standard deviation over five independent random draws of $k$.}
\label{tab:res2}
\end{table*}

\subsection{Scenario 1: LLM only}
Our proposed MARBLE-LLM consistently improves out-of-distribution generalization, particularly in the low-data regime (Table \ref{tab:res1} and Fig.~\ref{fig:res_bubbles}). For example, at $k$ = 100, our method achieves a balanced accuracy of 62.7$\pm$0.8, substantially outperforming both the supervised baseline (55.7$\pm$0.6) and the multi-hot pretraining approach (57.0$\pm$1.1). This advantage persists across larger sample sizes, where MARBLE-LLM continues to outperform alternatives in the out-of-distribution setting. The gains of MARBLE-LLM compared to the multi-hot and summarization approaches across all evaluation settings demonstrate the benefits of using embeddings and of learning to aggregate biomarker embeddings dynamically, respectively.
Strong performance on in-distribution biomarkers is also important to consider, as it demonstrates that strong out-of-distribution generalization does not come at a cost of in-distribution performance. MARBLE-LLM also almost always has the best in-distribution performance as well. The only setting where another approach has better balanced accuracy is for the supervised model when $k=100$ and on in-distribution biomarkers (67.4$\pm$0.4 versus 66.6$\pm$0.3 for supervised and MARBLE-LLM, respectively). However, because of the large gains of MARBLE-LLM on out-of-distribution biomarkers, the overall balanced accuracy of MARBLE-LLM is still higher. 

\subsection{Scenario 2: All settings}
In the second scenario, we focus on the relative value of the different biomarker representations.
Table \ref{tab:res2} and Fig.~\ref{fig:res_bubbles} present the results of the three pretraining settings. We observe that MARBLE-LLM pretraining provides stronger out-of-distribution generalization than \mbox{MARBLE-PLM} across all data regimes. For example, in the low-data regime at $k$ = 100, LLM-based embeddings achieve a BA of 62.7$\pm$0.5, compared to 61.3$\pm$0.1 with PLM-based embeddings.

When combining both PLM and LLM embeddings, MARBLE achieves consistently strong performance, often matching or surpassing the best individual approach. For example, at 
$k$ = 10000, the combined model achieves a performance of 68.6$\pm$0.2, the highest overall and comparable to MARBLE-LLM in Scenario 1 despite less pretraining data. 
These findings suggest that LLM embeddings are particularly effective for enhancing generalizability, while their integration with PLM embeddings yields complementary benefits that improve performance across both in- and out-of-distribution scenarios.

Comparison across Tables~\ref{tab:res1} and \ref{tab:res2} for MABLE-LLM allows us to consider the impact of the size of the pretraining dataset. Generally, we observe a slight boost in performance with the larger pretraining dataset when considering overall balanced accuracy, however it is small or even zero for out-of-distribution biomarkers.
For example, at $k$ = 100 in the out-of-distribution setup, LLM-based pretraining achieves 62.7$\pm$0.8 on the full dataset and 
62.7$\pm$0.5 on the restricted subset. This observation highlights the robustness of LLM-based embeddings and suggests that the advantage in low-data regime fine-tuning can be realized even with only modest amounts of pretraining data. At the same time, using the full dataset provides a slight performance boost at larger $k$ (e.g., 69.3$\pm$0.2 versus 68.2$\pm$0.2 at $k$ = 10000), indicating that LLM-based pretraining can still benefit from a larger pretraining cohort.

\subsection{Statistical significance of per-biomarker performance improvements}
While the above results show the value of MARBLE at the panel-level, we also wanted to understand how these improvements are distributed among the different genomic biomarkers. To do so, we conducted a paired one-sided t-test on a per-biomarker basis (Table~\ref{tab:res_stattest}). For each biomarker, balanced accuracy across five independent draws of $k$ is compared between the corresponding pair of models. Overall, these results suggest that MARBLE and its variants improve performance for many biomarkers with a minimal number of biomarkers exhibiting a degradation of performance (Fig.~\ref{fig:delta} for detailed deltas). For instance, at \mbox{$k=100$}, 47.2\% of out-of-distribution biomarkers demonstrated significantly higher balanced accuracy with \mbox{MARBLE} ($p<0.05$), while only 1.9\% of biomarkers showed significantly lower performance compared to the supervised baseline. The remaining biomarkers exhibited no significant difference between the models. Across larger fine-tuning sets ($k=1000$ and $k=10000$), the proportion of biomarkers with significantly higher performance for the MARBLE and its variants decreased, reflecting smaller gains as more labeled data became available. Comparison between \mbox{MARBLE-LLM}, \mbox{MARBLE-PLM}, and \mbox{MARBLE} revealed complementary benefits, with each contributing significant gains for different subsets of biomarkers (Table~\ref{tab:res_stattest}). This statistical analysis confirms that multimodal pretraining yields the largest improvements when labeled data is scarce.

\begin{table}[ht!]
\centering
\small
\resizebox{0.47\textwidth}{!}{
{\renewcommand{\arraystretch}{1.15} 
\begin{tabular}{lcc|cc|cc}
\toprule
                          & \multicolumn{2}{c|}{k=100}   & \multicolumn{2}{c|}{k=1000}  & \multicolumn{2}{c}{k=10000}  \\ \hline
                          & \textgreater{} & \textless{} & \textgreater{} & \textless{} & \textgreater{} & \textless{} \\ \hline
MARBLE-LLM vs. supervised & 52.8           & 3.8         & 26.4           & 1.9         & 30.2           & 11.3        \\
MARBLE-PLM vs. supervised & 41.5           & 3.8         & 34.0           & 13.2        & 17.0           & 7.5         \\
MARBLE vs. supervised & 47.2           & 1.9         & 39.6           & 5.7        & 28.3           & 11.3         \\
MARBLE-LLM vs. MARBLE-PLM & 20.8           & 9.4         & 18.9           & 11.3        & 9.4            & 7.5         \\
MARBLE-LLM vs. MARBLE     & 13.2           & 13.2        & 13.2           & 13.2        & 1.9            & 3.8         \\
MARBLE-PLM vs. MARBLE     & 3.8            & 13.2        & 5.7            & 11.3        & 5.7            & 5.7         \\ \bottomrule
\end{tabular}
}}
\caption{\textbf{Per-biomarker statistical significance analysis of model comparisons.} Each cell reports the percentage of biomarkers for which the first model achieved significantly higher (\textless{}) or lower (\textgreater{}) balanced accuracy than the second model across different fine-tuning data sizes ($k = 100, 1000, 10000$).}
\label{tab:res_stattest}
\end{table}
\section{Conclusion}
In this work, we present MARBLE, a multimodal contrastive pretraining framework that aligns histopathology WSIs with biomarker representations derived from LLMs and PLMs. Our results demonstrate that integrating biomarker knowledge through contrastive pretraining substantially improves out-of-distribution generalization, especially in low-data regimes and without degradation of in-distribution performance. LLM-based embeddings proved especially effective for generalization, and their integration with PLM embeddings provided additional improvement. 
These findings highlight the potential of combining pathology FMs with structured biological knowledge. This integration is crucial for precision oncology applications, particularly as new genomic or molecular biomarkers are discovered and incorporated into sequencing panels. 
While our study focused on leveraging protein-level and text-based biomarker representations, future work could incorporate other representations such as from recently developed DNA foundation models \cite{nguyen2024sequence,brixi2025genome}. Overall, our proposed approach offers encouraging evidence for the expanding use of H\&E images in precision oncology applications.

\normalsize
\bibliography{main}

\footnotesize
\newpage
\title{\Large Supplementary Information}
\clearpage
\setcounter{page}{1}
\renewcommand{\thesection}{S\arabic{section}}
\setcounter{section}{0}
\renewcommand{\thefigure}{S\arabic{figure}}
\renewcommand{\thetable}{S\arabic{table}}
\setcounter{figure}{0}
\setcounter{table}{0}

\section{Additional model details}
\subsection{Histopathology tile embedding model}
We used the CLS token from Virchow2~\cite{zimmermann2024virchow2} as recommended by the authors, which produces tile embeddings with a dimensionality of $d_H=1280$.

\subsection{Aggregator for tile embeddings}
\label{sec:supp_agg_emb}
The keys and values in the Agata pathology features aggregator module are obtained from a set of tile embeddings $\textbf{H}_{i}$ from a single slide using learned transformations:
\begin{align}
    \textbf{K}_{i} &= \text{GELU}(\textbf{H}_i\textbf{W}_{1} + \textbf{1}_{N_H^i}\textbf{c}_{1}), \\
    \textbf{V}_{i} &= \text{GELU}(\textbf{K}_{i}\textbf{W}_{2} + \textbf{1}_{N_H^i}\textbf{c}_{2})
\end{align}
where $\textbf{W}_{1} \in \mathbb{R}^{d_H  \times d_{k}}, \textbf{W}_{2} \in \mathbb{R}^{d_{k}  \times d_{v}}$ and $\textbf{c}_{1} \in \mathbb{R}^{d_{k}}, \textbf{c}_{2} \in \mathbb{R}^{d_{v}}$ are learnable parameters and $\textbf{1}_{N_H^i}$ is an $N_H^i$-dimensional vector of ones. In our implementation, the input embedding dimension is $d_H$=1280, and the two feed-forward layers have output dimensions $d_{k}$=320 and $d_{v}$=640. $\textbf{K}_i$ and $\textbf{V}_i$ are used in the cross-attention mechanism as described in the text along with the learnable queries $\textbf{Q}\in\mathbb{R}^{M \times d_{v}}$, where $M=16$.

\subsection{Biomarker embedding models}
We used all-MiniLM-L6-v2, a lightweight 6-layer Transformer model from the SentenceTransformers library, which produces sentence embeddings with a dimensionality of $d_{b_{l}} = 384$, optimized for semantic similarity and retrieval tasks.

For PLM-based embeddings, we used the pretrained ESM-2 model, which produces representations of altered amino acid sequences derived from each variant with a dimensionality of $d_{b_{p}} = 1280$.

\subsection{Aggregator for biomarker embeddings derived by LLM}
\label{sec:supp_agg_biom_emb_llm}
For biomarkers embedded using LLM, the multi-head cross-attention aggregator operates as follows:
\begin{equation}
    \begin{aligned}
        &\textbf{Q}_{L}=\textbf{q}_{L}\textbf{W}_{Q,L}^{(head)}, \; \textbf{K}_{L,i}=\textbf{b}_{L,i}\textbf{W}_{K,L}^{(head)}, \; \textbf{V}_{L,i}=\textbf{b}_{L,i}\textbf{W}_{V,L}^{(head)}, \\
        &\text{Attn}_{L,head}(\textbf{Q}_{L},\textbf{K}_{L,i},\textbf{V}_{L,i}) = \text{softmax}\!\left(\frac{\textbf{Q}_{L}\textbf{K}_{L,i}^{\top}}{\sqrt{d_{b_{L}}}}\right)\textbf{V}_{L,i}, \\
        &\textbf{b}_{\text{agg,L,i}} = \text{Concat}_{head} \big[ \text{Attn}_{L,head}(\textbf{Q}_{L},\textbf{K}_{L,i},\textbf{V}_{L,i}) \big] \textbf{W}_{O,L}.
    \end{aligned}
\end{equation}
where $\textbf{Q}_{L}\in\mathbb{R}^{M \times d_{b_{L}}}$, $\textbf{K}_{L,i}\in\mathbb{R}^{N_L^i \times d_{b_{L}}}$, and $\textbf{V}_{L,i} \in \mathbb{R}^{N_L^i \times d_{b}}$ are query, key, and value matrices, respectively, $M=1$ is the number of learned queries. The projection matrices are $\textbf{W}_{Q,L}^{(head)}\in\mathbb{R}^{d_{b_{L}} \times d_{b_{L}}}$, $\textbf{W}_{K,L}^{(head)}\in\mathbb{R}^{d_{b_{L}} \times d_{b_{L}}}$, $\textbf{W}_V^{(head)}\in\mathbb{R}^{d_{b_{L}} \times d_{b_{L}}}$, and $\textbf{W}_{O,L}\in\mathbb{R}^{(H d_{b_{L}}) \times d_{b_{L}}}$, where $H=8$ is the number of attention heads.
$\textbf{b}_{L,i}$ is a set of LLM-based biomarker embeddings from  $\textbf{L}_i$.

\subsection{Aggregator for biomarker embeddings derived by PLM}
\label{sec:supp_agg_biom_emb_plm}
For biomarkers embedded using PLM, an analogous cross-attention aggregator is applied:
\begin{equation}
    \begin{aligned}
        &\textbf{Q}_{P}=\textbf{q}_{P}\textbf{W}_{Q,P}^{(head)}, \; \textbf{K}_{P,i}=\textbf{b}_{P,i}\textbf{W}_{K,P}^{(head)}, \; \textbf{V}_{P,i}=\textbf{b}_{P,i}\textbf{W}_{V,P}^{(head)}, \\
        &\text{Attn}_{P,head}(\textbf{Q}_{P},\textbf{K}_{P,i},\textbf{V}_{P,i}) = \text{softmax}\!\left(\frac{\textbf{Q}_{P}\textbf{K}_{P,i}^{\top}}{\sqrt{d_{b_{P}}}}\right)\textbf{V}_{P,i}, \\
        &\textbf{b}_{\text{agg,P,i}} = \text{Concat}_{head} \big[ \text{Attn}_{P,head}(\textbf{Q}_{P},\textbf{K}_{P,i},\textbf{V}_{P,i}) \big] \textbf{W}_{O,P}.
    \end{aligned}
\end{equation}
where $\textbf{Q}_{P}\in\mathbb{R}^{M \times d_{b_{P}}}$, $\textbf{K}_{P,i}\in\mathbb{R}^{N_L^i \times d_{b_{P}}}$, and $\textbf{V}_{P,i} \in \mathbb{R}^{N_P^i \times d_{b}}$ are query, key, and value matrices, respectively, $M=1$ is the number of learned queries. The projection matrices are $\textbf{W}_{Q,P}^{(head)}\in\mathbb{R}^{d_{b_{P}} \times d_{b_{P}}}$, $\textbf{W}_{K,P}^{(head)}\in\mathbb{R}^{d_{b_{P}} \times d_{b_{P}}}$, $\textbf{W}_V^{(head)}\in\mathbb{R}^{d_{b_{P}} \times d_{b_{P}}}$, and $\textbf{W}_{O,P}\in\mathbb{R}^{(H d_{b_{P}}) \times d_{b_{P}}}$, where $H=8$ is the number of attention heads.
$\textbf{b}_{P,i}$ is a set of LLM-based biomarker embeddings from  $\textbf{P}_i$.

\begin{figure*}[]
    \centering
    \includegraphics[width=0.8\linewidth]{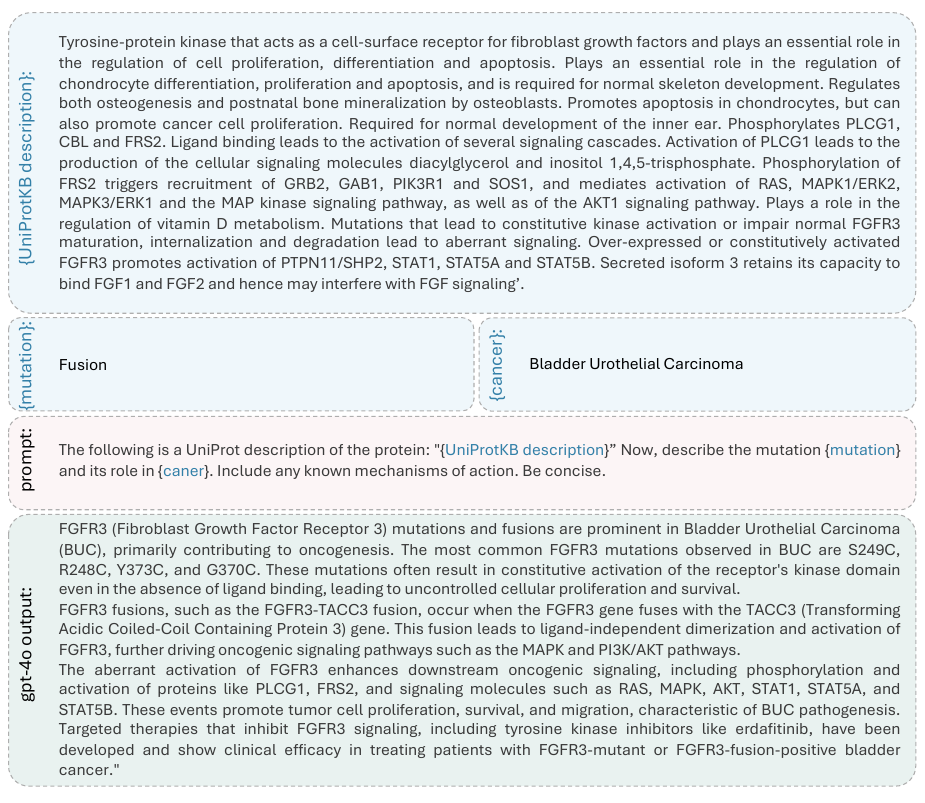}
    \caption{\textbf{Example of the biomarker-to-text transformation.} 
For each biomarker defined by an oncogene, mutation, and cancer type, a canonical protein description is retrieved from UniProtKB and provided to an LLM (gpt-4o), which generates a concise text paragraph summarizing the mutation's role 
and the mechanism of action in the specified cancer. The resulting output (e.g., shown here for FGFR3 fusion in bladder urothelial carcinoma) is then embedded into a fixed-length vector representation using a text embedding model.}
\label{fig:prompt}
\end{figure*}

\begin{figure*}[!t]
    \centering
    \includegraphics[width=0.8\linewidth]{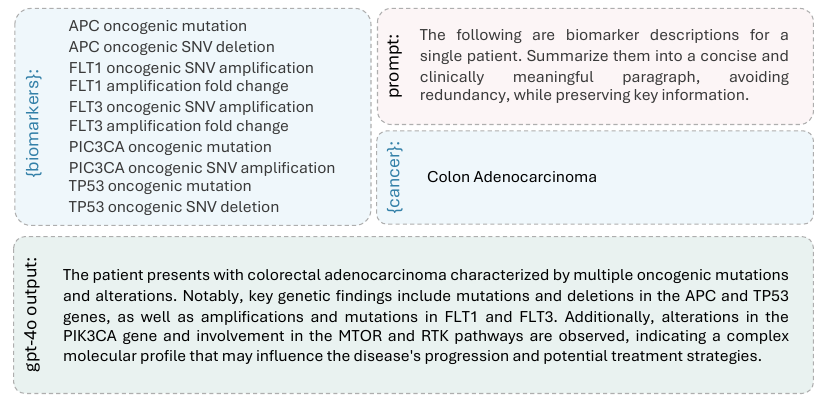}
    \caption{\textbf{Example of the biomarker-to-text transformation for LLM-based summarization approach.} 
The LLM is first prompted to generate a concise paragraph describing all biomarkers for a given patient. The resulting output is then embedded into a fixed-length vector representation using a text embedding model.}
\label{fig:prompt_summary}
\end{figure*}

\begin{table}[]
\centering
\begin{tabular}{lc}
Parameter   & Value  \\ \hline
Model       & gpt-4o \\
Temperature & 1      \\ \hline
\end{tabular}
\caption{\textbf{API parameters for GPT-4 biomarker description generation.}}
\label{tab1S}
\end{table}

\section{Experimental Details}
\label{sec:suppl_impl_det}
\subsection{Contrastive pretraining}
Before aggregation, each biomarker embedding is passed through a small MLP (Linear–ReLU–Linear, $d\rightarrow 2d\rightarrow d$). A one-hot metadata vector encoding (i) panel version and (ii) primary vs.\ metastatic status is projected through a linear layer and appended to the biomarker set. This produces an input of shape $[B, N_L{+}2, d_{b_{L}}]$ for LLM-based biomarker embeddings and $[B, N_P{+}2, d_{b_{P}}]$ for PLM-based embeddings to the cross-attention aggregator, where $B$ is the batch size, $N_L$ is the number of LLM-based biomarker embeddings for a given patient, and $N_P$ is the number of PLM-based biomarker embeddings for a given patient.
The outputs of the tile-level and biomarker-level aggregators are each fed into modality-specific projection heads for contrastive learning, each implemented as a two-layer multilayer perceptron with residual connection and layer normalization. Concretely, each projection head linearly maps the aggregated embedding to a 256-dimensional space, applies GELU activation, a second linear transformation with dropout, and a residual skip connection before layer normalization.

\begin{figure*}[]
    \centering
    \includegraphics[width=1.0\linewidth]{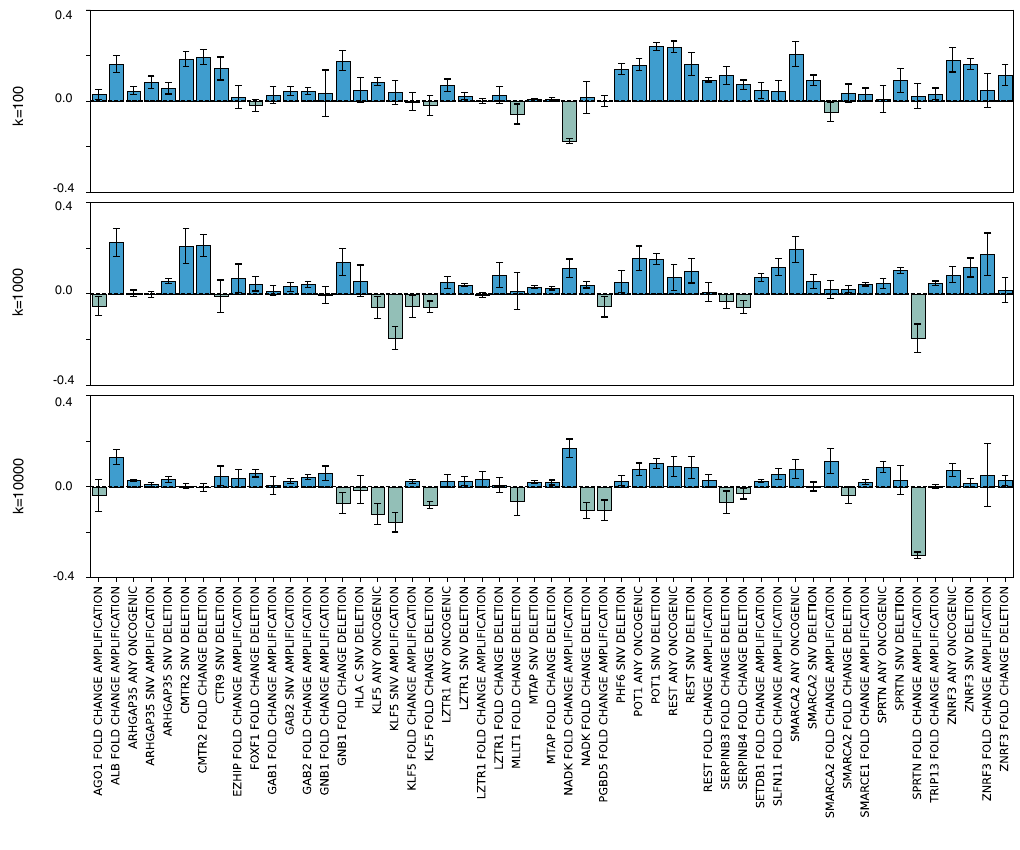}
    \caption{\textbf{Per-biomarker performance difference between MARBLE and supervised model.} Difference shown for out-of-distribution biomarkers. The error bars $+/-$ one standard deviation for the five trials.}
\label{fig:delta}
\end{figure*}

\end{document}